\documentclass[11pt,a4paper]{scrartcl}

\usepackage{CLICdp}

\usepackage{CLICdp_definitions}
\usepackage{pst-pdf}
\usepackage[symbol]{footmisc}
\usepackage{feynmf}
\usepackage{gmp}
\usepackage{textpos}
\usepackage{paralist}
\usepackage{xcolor}

\title{Measurement of Higgs decay to WW* \\
		in Higgsstrahlung at $\sqrt{s}$=500 GeV ILC and\\
                 in WW-fusion at $\sqrt{s}$=3 TeV CLIC}

\clicdpconf{2017}{007}  % conference proceedings
\date{\today}
\addauthor{Mila Pandurovi\'{c}}{\institute{1}\footnote{milap@vinca.rs}}

\addinstitute{1}{Vinca Institute of Nuclear Sciences, University of Belgrade, Serbia}

\onbehalfof{On behalf of the CLICdp collaboration and the ILC Physics and Detector Study}
\abstract{This talk presents results of the two independent analyses evaluating the measurement accuracy of the branching ratio for the Standard model Higgs boson decay to a W-pair, at the Compact Linear Collider (CLIC) and at the International Linear Collider (ILC). The considered Higgs production channels are the WW-fusion for the highest energy stage of CLIC, $\sqrt{s}$= 3 TeV, and the Higgsstrahlung process for the nominal ILC energy, $\sqrt{s}$=500 GeV. Both studies are performed using the full simulation of the detector. The realistic experimental conditions have been simulated including beam energy spectrum, initial state radiation and the backround from $\gamma\gamma\rightarrow hadrons$ processes, which are overlaid on simulated events. The multivariate analysis technique is used for the final event selection and the expected relative statistical uncertainty, $\Delta (\sigma \cdot BR) / (\sigma\cdot BR)$, of the measured Higgs production cross sections is estimated.}

\titlecomment{Talk presented at the International Workshop on Future Linear Colliders (LCWS16), Morioka, Japan, 6--10 December 2016. C16-12-05.4.}
\graphicspath{ {./logos/}{./figures/} }

\begin{document}
%\linenumbers
% generates the title page
\titlepage
\section{Introduction}

The important part of the physics program on the future linear electron-positron colliders (LC) are the precise measurements of the Higgs boson properties. The measurements of the Higgs boson couplings, for which the Standard model gives strict predictions, namely the linear dependence on the masses of corresponding particles, are one of the top priorities of the LC Higgs program. The shape of the possible deviations from these predictions depends on the proposed model of the new physics and the precision of the coupling measurements of the order of few percent is needed to be sensitive to these effects, if no other state related to electroweak symmetry breaking is directly accessible at the Large hadron collider \cite{JamesWells}. This sensitivity can be successfully achieved at the proposed future linear $e^{+}e^{-}$ collliders, which are best suited for precision measurements. \\

In the first part of this contribution the measurement of the Higgs decay into a pair of W bosons is considered, at the nominal center-of-mass energy, $\sqrt{s}$ = 500 GeV,  of ILC, using Higgsstrahlung as the Higgs production process. The relative statistical accuracy of the measurement of $\sigma (HZ) \times BR (H \rightarrow WW ^ {*}) $ have been estimated. The measured cross section is proportional to the coupling product ${g_{HZZ}^2\cdot g_{HWW}^2}/{\Gamma}_H$.

The second part of this contribution is dedicated to the same Higgs decay, $H \rightarrow WW ^ {*}$, but analyzed at the highest energy stage of CLIC, $\sqrt{s}$= 3 TeV, where the dominant Higgs production channel is the WW-fusion. The relative statistical uncertainty of the partial cross-section $\sigma (H\nu_{e}\nu_{e}) \times BR (H \rightarrow WW ^ {*}) $, is determined.

\section{Simulation and analysis tools}

Both analyses are using ILCSoft, a common software packages developed for the International Linear Collider.
Signal and background samples are simulated using the Whizard 1.95 \cite{WHIZARD} event generator, including initial state radiation and a realistic ILC or CLIC luminosity spectrum. 
The luminosity spectrum and beam-induced processes were simulated by GuineaPig 1.4.4 \cite{GUINEAPIG}. 
The hadronization and fragmentation of the Higgs and vector bosons are simulated using Pythia 6.4 \cite{Pythia}. 
Background coming from $\gamma\gamma$ to hadrons were overlaid over each generated event sample before reconstruction. 
Particle reconstruction and identification was done using the particle flow technique, implemented in the Pandora particle-flow algorithm (PFA) \cite{PFA1,PFA2}. 
The response of the detector was simulated with the CLIC$\_$ILD for CLIC and the ILD$\_$o1$\_$v05 detector model for ILC.
Signal and background separation is obtained using multivariate classification analysis, implemented in the TMVA package \cite{TMVA}.

For the ILC analysis the Higgs mass of $m_{H}$ = 125 GeV is assumed and an integrated luminosity of 500 fb$^{-1}$. Also, polarization of both, electron and positron, beams P(e$^-$, e$^{+}$) = ($-$80$\%$, +30$\%$). 
The CLIC analysis assumes $m_{H}$ = 126 GeV, an integrated luminosity of 2 ab$^{-1}$ and unpolarized beams.

\section{Higgs$\rightarrow$WW* in Higgsstrahlung at 500 GeV ILC}

At the nominal energy of the ILC, $\sqrt{s}$=500 GeV, and the considered beam polarizations the cross-section of the Higgsstrahlung process is 114 fb. For signal events the fully hadronic channel is considered, where the Z boson, as well as both W bosons coming from the Higgs decay, decay to quark pairs (six jet final state). The corresponding signal cross section is 11.33 fb. 
The Feynman diagram of the Higgsstrahlung Higgs production channel is shown in Figure \ref{f1}.

\begin{figure}[h]
\begin{minipage}[b]{0.5\linewidth}
\centering
\begin{fmffile}{AbreBrate}
\begin{center}
\begin{fmfgraph*}(140,80) 
\fmfleft{i1,i2}
\fmfright{o1,o2}

\fmf{fermion}{i1,v1} \fmf{fermion}{v1,i2}
\fmf{dashes}{v2,o1} \fmf{boson}{o2,v2}
\fmflabel{$e^-$}{i1}
\fmflabel{$e^+$}{i2}
\fmflabel{$H$}{o1}
\fmflabel{$Z^0$}{o2}

\fmfdot{v1,v2}
\fmf{photon,label=$Z^0$}{v1,v2}
\end{fmfgraph*}
\end{center}
\end{fmffile}
\caption{\label{f1} Higgsstrahlung Higgs production process.}
\end{minipage}    
\begin{minipage}[b]{0.5\linewidth}
\centering
\begin{fmffile}{wwfusion}
\begin{center}
\begin{fmfgraph*}(140,80) 
\fmfleft{i1,i2} \fmfright{o1,o2,o3}
\fmf{fermion,label=$e^-$}{i1,v1} \fmf{fermion,label=$e^+$,label.side=left}{v3,i2}
\fmf{fermion}{v1,o1} \fmf{fermion}{o3,v3}
\fmf{boson,label=$W^-$}{v1,v2} \fmf{boson,label=$W^+$,label.side=right}{v3,v2} 
\fmf{dashes,label=$H$}{o2,v2}
%\fmf{boson}{v3,o3} 
\fmfdot{v1,v2,v3}
%\fmflabel{$W$}{o2} \fmflabel{$W^*$}{o3}
\fmflabel{$\PGne$}{o1} \fmflabel{$\PAGne$}{o3}
\end{fmfgraph*}
\end{center}
\end{fmffile}
\caption{\label{f2} WW-fusion Higgs production process.}
\end{minipage}     
\end{figure}

\subsection {Background processes}

The background processes that are considered in this study are listed in Table \ref{Table:Background1}. 

\begin{table}[t]
\caption {List of the considered background processes, with the corresponding cross sections for $\sqrt{s}$ = 500 GeV and integrated luminosity of 500 fb$^{-1}$. The table lists the signal and background selection efficiencies, after the preselection and the final selection, and the expected number of events in the final sample.} \label{Table:Background1} 
\centering
\begin{tabular}{lrrrr}

\\\hline
 
$Process$                                       & $\sigma[fb]$  	& $\epsilon_{pres}$ [$\%$]& $\epsilon_{total}$ [$\%$]& $evts$$_{final}$\\
\hline
signal						&	   11.3	        & 79.7 	& 22.68 &1285\\
\hline
$H\rightarrow$ other Higgs decays         	&  	  103.4       	& 59.4	& 3.35	&1733 \\ 
$e^{+}e^{-}\rightarrow $ 4f ZZ hadronic		&         680.2       	& 44.0	& 0.07	&226\\
$e^{+}e^{-}\rightarrow $ 4f WW hadronic		&  	 7680.7   	& 29.1	& 0.17	&616\\
$e^{+}e^{-}\rightarrow $ 4f WW/ZZ mixhadronic   &  	 6400.1		& 29.4	& 0.01	&431\\
$e^{+}e^{-}\rightarrow $ 4f ZZ semileptonic     &  	  608.6		& 4.5  &$\textless 10^{-4}$&-\\
$e^{+}e^{-}\rightarrow $ 4f WW semileptonic   	&  	 9521.4        	& 2.0  & 1$\cdot10^{-4}$&27\\ 
$e^{+}e^{-}\rightarrow $ 2f hadronic   		& 	32470.5        	& 3.4	&0.01	&1683\\
$e^{+}e^{-}\rightarrow $ 6f $t\bar{t}$ yyxyev	&  	  116.9       	& 44.4	&0.12	&106\\
$e^{+}e^{-}\rightarrow $ 6f $t\bar{t}$ yyveyx  	&  	  117.1        	& 44.5	&0.14	&114\\
$e^{+}e^{-}\rightarrow $ 6f $t\bar{t}$ yyuyyc	&  	  164.4        	& 44.1	&0.13	&107\\
$e^{+}e^{-}\rightarrow $ 6f $t\bar{t}$ yycyyu	&  	  165.5        	& 44.5	&0.12	&103\\
$e^{+}e^{-}\rightarrow $ 6f $t\bar{t}$ yyxylv  	&  	  231.1        	& 53.9	&0.12	&73\\
$e^{+}e^{-}\rightarrow $ 6f $t\bar{t}$ yyvlyx  	&  	  231.6        	& 54.0	&0.12	&69\\
$e^{+}e^{-}\rightarrow $ 6f $t\bar{t}$ yyucuuc	&  	  163.3        	& 58.2	&0.14	&161\\
$e^{+}e^{-}\rightarrow $ 6f $t\bar{t}$ yyuyyu	&  	  166.6        	& 58.2	&0.15	&174\\

 \hline
\end{tabular}
\end{table}

\section{Event selection}

\begin{figure}[t]

\begin{minipage}[b]{0.5\linewidth}
\includegraphics[width=7.5cm]{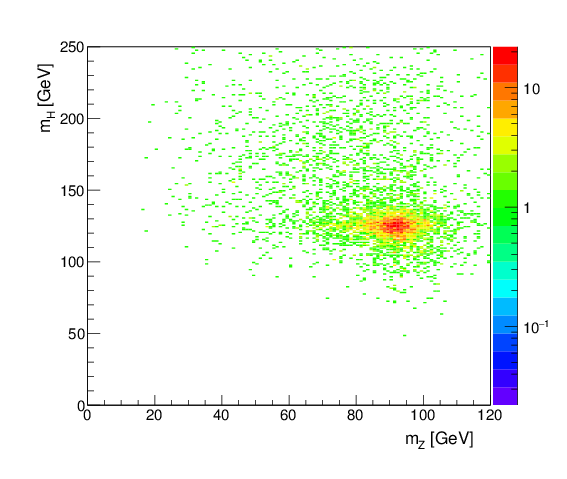}
\end{minipage}
\begin{minipage}[b]{0.5\linewidth}
\includegraphics[width=7.5cm]{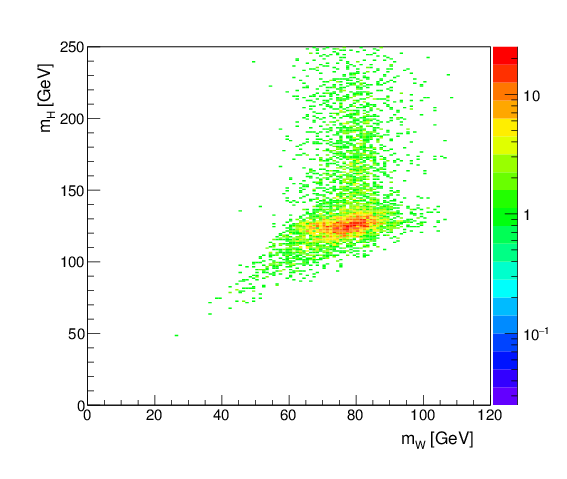}

\end{minipage}
  \caption{The result of the jet pairing. Left: the invariant mass of the Z boson candidate (m$_Z$) vs. the invariant mass of the Higgs boson candidate (m$_H$). Right: the invariant mass of the real W boson candiate m$_W$ vs. the invariant mass of the Higgs boson candiate m$_H$ (right).}
    \label{fig:JetPairMass}
\end{figure}

Event selection is performed in several steps.  First, all reconstructed particles are clustered into six jets using the $k_{T}$ clustering algorithm. The b and c-tagging probabilities, determined by LCFIPlus package, are assigned to each jet in the event. In the next step, the signal process kinematics is reconstructed by pairing of jets to form candidate for the Z boson, as well as, one on-shell and one off-shell W boson, coming from Higgs decay. The combination of the jet pairs is chosen by the minimization of the $\chi^{2}$ function given by the formula:

$$\chi^{2}= \frac{m_{ij}-m_{W}} {{\sigma_{W}}^{2}}+ \frac{m_{kl}-m_{Z}} {{\sigma_{Z}}^{2}}+\frac{m_{ijmn}-m_{H}} {{\sigma_{H}}^{2}}   $$

\noindent where the invariant mass of a di-jet pair m$_{ij}$ is assigned to the candidate for the real W boson, m$_{kl}$ is assigned to the Z boson candidate, while  m$_{ijmn}$ is the invariant mass of the Higgs boson candidate. m$_{V}$ and $\sigma_{V}$, (V = W, Z, H), are the masses and the expected mass resolutions of the corresponding bosons. The illustration of the jet pairning  is given in Figure \ref{fig:JetPairMass}.

The cross sections of the considered background processes are several orders of magnitude higher than the signal cross section (see Table \ref{Table:Background1}), therefore at the next step, the background to signal ratio is minimized by the set of preselection criteria prior to the final selection. The variables with the corresponding cut-off values used in the preselection, are:   
\begin{itemize}
  \item the invariant mass of the Z boson candidate, $70< m_{Z} < 110$ GeV;
  \item number of particle flow objects, NPFO$>$40;
  \item event thrust$>$0.95;  
  \item -log(y$_{45}) < 4.4 $;
  \item -log(y$_{56}) < 4.8$;
\end{itemize}

\noindent where y$_{ij}$ is the value of the $k_T$ algorithm parameter at which the number of reconstructed jets changes from i to j.

Efficiencies of the preselection are given in Table \ref{Table:Background1}, for signal and background processes.

The final event selection is based on the multivariate analysis method using the Boosted decision tree (BDT) algorithm. It exploits kinematic properties of the event in order to reject the residual background contribution. All background processes are used in the training of the algorithm. The list of discriminating input variables include:
\begin{itemize}
\item invariant masses of both W, Z and Higgs bosons, $m_{W}$, $m_{W^{*}}$, $m_{Z}$, $m_{H}$;
\item number of particle-flow objects (NPFO) in the event;
\item total visible energy, $E_{vis}$;
\item transverse momentum of jets that comprize the Higgs boson, $p_{T}^{Higgs}$;
\item jet reonstruction parameters -log($y_{12}$), -log($y_{23}$), -log($y_{34}$), -log($y_{45}$), -log($y_{56}$), -log($y_{67}$);
\item event shape variables (thrust, oblateness, sphericity and aplanarity);
\item and flavor tagging probabilities for the six reconstructed jets, btag$_i$, ctag$_i$ (i=1,6).
\end{itemize}

A cut-off value on the output of the BDT algorithm is used for the final separation of signal and background events and it is optimized to minimize the ratio: 

\begin{equation}
\frac{\Delta\sigma}{\sigma} =\frac{N_{S}}{\sqrt{(N_{S}+N_{B})}},
\end{equation}

\noindent where $N_{S}$, $N_{B}$ are the number of signal and background events after the final selection, respectively.\\
After the final selection the dominant backgrounds come from other Higgs decays, due to the kinematical similarity, as well as, from $q\bar{q}q\bar{q}$ and $q\bar{q}$ processes (see Table \ref{Table:Background1} ) due to the very high cross-sections.
The obtained relative statistical uncertainty on the product of the Higgsstrahlung cross-section and the corresponding branching ratio, $\sigma(HZ)\times BR(H\rightarrow WW^{*})$, is 6.5$\%$ at the 500 GeV ILC, assuming integrated luminosity of 0.5 ab$^{-1}$.

\begin{table}[t]
\caption {List of considered background processes, with the corresponding cross-sections and number of events for $\sqrt{s}$ = 3 TeV and assumed integrated luminosity of 2 ab$^{-1}$. The table also lists the signal and background reduction efficiencies, after the preselection and the final selection. The last column gives the expected number of events in the final sample. The beamstrahlung photons are denoted BS, while the phtons from equivalent photon approximation are denoted as EPA.} \label{Table:Background2} 
\centering
\begin{tabular}{lrrrr}

\\\hline
 
$Process$                                 & $\sigma[fb]$&   $\epsilon_{pres}$ [$\%$]& $\epsilon_{total}$ [$\%$]&  $evts$$_{final}$ \\
\hline

$H\rightarrow$ other Higgs decays  	&    374.3     	    &	64.6	&	18.0				&14534\\ 

$e^{+}e^{-}\rightarrow $$q\bar{q}$	   	&   2948.9  	    &	2.0	&	6$\cdot10^{-4}$	&38\\
$e^{+}e^{-}\rightarrow $$q\bar{q}\nu\nu$		&   1317.5	    &	45.8	&	0.3				&7664\\
$e^{+}e^{-}\rightarrow $$q\bar{q}l\nu$		&   5561.1  	    &	26.3	&	0.1				&12623\\
$e^{+}e^{-}\rightarrow $$q\bar{q}ll$		&   3319.6	    &	4.0	&	0.1				&135\\
\hline	
$e^{+}e^{-}\rightarrow $$q\bar{q}q\bar{q}$		&    546.5  	    &	3.3	&	7$\cdot 10^{-2}$		&77\\
$e^{+}e^{-}\rightarrow $$q\bar{q}qqvv$		&     71.5	    &	2.2	&	0.3				&358\\
$e^{+}e^{-}\rightarrow $$q\bar{q}q\bar{q}l\nu$		&    106.9	    &	1.1	&	0.04			&93\\
$e^{+}e^{-}\rightarrow $$q\bar{q}q\bar{q}ll$		&    169.3	    &	1.8	&	0.05			&172\\
\hline		
$e^{+}e^{-}\rightarrow $$q\bar{q}q\bar{q}e$ (EPA)	&     54.2 	    &	2.1	&	0.15			&161\\
$e^{+}e^{-}\rightarrow $$q\bar{q}q\bar{q}e$ (BS)	&    262.5	    &	3.3	&	$\textless10^{-4}$	&-\\
$e^{+}e^{-}\rightarrow $$q\bar{q}q\bar{q}e$ (EPA)	&     54.2  	    &	2.2	&	0.14			&146\\
$e^{+}e^{-}\rightarrow $$q\bar{q}q\bar{q}e$ (BS)	&    262.3 	    &	3.2	&	8$\cdot10^{-4}$		&4\\
\hline	
$e^{\pm}\gamma\rightarrow $$q\bar{q}q\bar{q}\nu$ (EPA)	&    287.8  	    &	2.0	&	0.05			&306\\
$e^{\pm}\gamma\rightarrow $$q\bar{q}q\bar{q}\nu$ (BS)	&   1268.6          &	2.6	&	0.04			&1082\\
$\gamma e^{\pm}\rightarrow $$q\bar{q}q\bar{q}\nu$ (EPA)	&    287.8	    &	2.2	&	0.07			&406\\
$\gamma e^{\pm}\rightarrow $$q\bar{q}q\bar{q}\nu$ (BS)	&   1267.3	    &	2.6	&	0.05			&1182\\
\hline	
$\gamma\gamma\rightarrow $$q\bar{q}q\bar{q}$ (EPA)(EPA)	&    402.7	    &	2.8	&	0.04		&368\\
$\gamma\gamma\rightarrow $$q\bar{q}q\bar{q}$ (EPA)(BS)	&   2423.1	    &	2.8	&	0.24		&1161\\
$\gamma\gamma\rightarrow $$q\bar{q}q\bar{q}$ (BS)(EPA)	&   2420.6	    &	2.7	&	0.34		&1659\\
$\gamma\gamma\rightarrow $$q\bar{q}q\bar{q}$ (BS)(BS)	&  13050.3	    &	2.0	&	4$\cdot10^{-4}$		&107\\
 \hline
\end{tabular}
\end{table}

\subsection{Higgs$\rightarrow$WW* in WW-fusion at 3 TeV CLIC}

The Higgs production at the highest CLIC energy stage, $\sqrt{s}$=3 TeV, is dominated by the WW-fusion process (see Figure \ref{f2} ). The boosted topology of this process is reflected in the signature of the signal: the Higgs decay studied is characterized by four soft, forward-peaked jets and the missing energy. The total invariant mass of jets in the event is consistent with the Higgs boson mass and the invariant mass of one of the jet pairs has to be consistent with the invariant mass of the W boson. The list of signal and considered background processes is given in Table \ref{Table:Background2} for the assumed integrated luminosity of 2 ab$^{-1}$.

\section{Event selection}

Events are clustered into four jets using the $k_{T}$ clustering algorithm. The opening of the jet cone was set to R=0.9, which gave the best invariant mass resolution for the Higgs and the real W boson, and the best mean invariant mass value. Jet are combined into pairs, and the combination, which gives the invariant mass of the jet pair closest to the mass of the real W boson, is chosen.

The following preselection cuts are applied to minimize the high cross section backgrounds:
\begin{itemize}
  \item the invariant mass of the H boson, $90< m_{H} < 150$ GeV;
  \item number of particle flow objects, $p_t>$40.
\end{itemize}

Efficiencies of the preselection are given in Table \ref{Table:Background2}, for signal and background processes. After the preselection the main backgrounds are qql$\nu$, qq$\nu\nu$ and other Higgs decay processes, mainly, H$\rightarrow b\bar{b}$, H$\rightarrow gg$.\\
The final event selection is again based on using the multivariate analysis method, using the Boosted decision tree (BDT) algorithm. All background are used in the BDT training. The list of discriminating input variables include:

\begin{itemize}
\item total visible energy, $E_{vis}$;
\item the invariant masses of Higgs, real W and virtual W$^*$ candidates, $m_{W}$, $m_{W^{*}}$, $m_{H}$;
\item number of particle-flow objects (NPFO) in the event;
\item transverse momentum of each reconstructed jet in the event, $p_{t}$;
\item jets reconstruction parameters, -log($y_{12}$), -log($y_{23}$), -log($y_{34}$), -log($y_{45}$), -log($y_{56}$);
\item event thrust;
\item flavor tagging probabilities for the two jet hypothesis, btag$_i$, ctag$_i$, i=1,2.
\item angle between jets that comprise real W boson.
\end{itemize}

\begin{figure}[h]
%    \centering
    \includegraphics[width=1.0\textwidth]{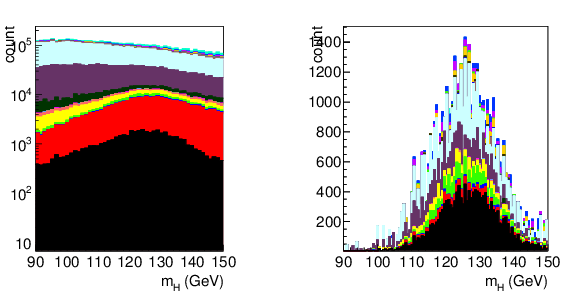}
  \begin{textblock}{1.}(10.6,-6.8)
    \textblockcolour{white}
  \rotatebox{0}{$\sqrt{s}$=3 TeV}
\end{textblock}
   \begin{textblock}{1.}(13.5,-6.8)
    \textbf{CLICdp}
    \end{textblock}
 
    \caption{Distribution of the reconstructed invariant mass for the Higgs boson candidate in the $H\rightarrow WW^{*}$ decay channel for signal (black) and backgrounds (colours) events after preselection (left) and final selection (right).}
    \label{fig:Stack3TeV}
\end{figure}

The final selection efficiencies are given in Table \ref{Table:Background2}. Figure \ref{fig:Stack3TeV} represents the stacked histogram of the signal (black) and background processes after the preselection (left) and after the final selection (right). The dominant backgrounds come from other Higgs decays, $H\rightarrow b\bar{b}$ (red), $H\rightarrow gg$ (light green), as well as, from $q\bar{q}\nu\nu$ (violet) and $q\bar{q}l\nu$ (light blue).

The relative statistical uncertainty of the measurement of $\sigma(H\nu_e\nu_e)\times BR(H\rightarrow WW^{*})$, expected at $\sqrt{s}$=3 TeV CLIC with the integrated luminosity of 2.0 ab$^{-1}$ is 1.5 $\%$.

\section{Conclusion}

Presented in this contribution are results of the two independent studies of cross section times branching fraction measurement for Higgs decaying to a W pair, at the ILC and  CLIC. Fully hadronic final states are considered. Both studies are are based on the full detector simulation, including initial state radiation and beam induced backgrounds. \\
The first study addresses the measurement at the nominal ILC energy, $\sqrt{s}$=500 GeV, using Higgsstrahlung Higgs production channel. The beam polarizations of P(e$^-$, e$^{+}$) = ($-$80$\%$, +30$\%$), the integrated luminosity of 500 fb$^{-1}$ and the mass of Higgs boson of 125 GeV, are assumed.  The obtained result for the relative statistical uncertainty of $\sigma(HZ)\cdot BR(H\rightarrow WW^*) $ is 6.5$\%$.\\
The second analysis is dedicated to the study of the $H\rightarrow WW^*$ decay at the highest energy stage of CLIC, $\sqrt{s}$=3 TeV, using the leading Higgs production channel, WW-fusion. The integrated luminosity of 2 ab$^{-1}$, the unpolarized beams and the mass of Higgs boson of 126 GeV are assumed. The obtained result for the relative statistical uncertainty of the $\sigma(H\nu_e\nu_e)\cdot BR(H\rightarrow WW^*) $ is 1.5$\%$.

% add references
\printbibliography[title=References]

\end{document}